\documentclass[aip,pop,floatfix,preprint]{revtex4-1}
\usepackage{amsmath}
\usepackage{amssymb}
\usepackage{graphicx}

\makeatletter

\usepackage{hyperref}
\usepackage{algorithm}
\usepackage{algpseudocode}
\usepackage{amsfonts}
\usepackage{subfigure}
\usepackage{multirow}
\usepackage{verbatim}
\usepackage{bm}

\makeatother

\begin{document}

\title{Reducing Noise for PIC Simulations Using Kernel Density Estimation
Algorithm}

\author{Wentao Wu}

\affiliation{School of Physics, University of Science and Technology of China,
Hefei, Anhui 230026, China}

\author{Hong Qin}

\affiliation{School of Physics, University of Science and Technology of China,
Hefei, Anhui 230026, China}

\affiliation{Plasma Physics Laboratory, Princeton University, Princeton, NJ 08543,
USA}

\thanks{Corresponding author, hongqin@ustc.edu.cn}
\begin{abstract}
Noise is a major concern for Particle-In-Cell (PIC) simulations. We
propose a new theoretical and algorithmic framework to evaluate and
reduce the noise level for PIC simulations based on the Kernel Density
Estimation (KDE) theory, which has been widely adopted in machine
learning and big data science. According to this framework, the error
on particle density estimation for PIC simulations can be characterized
by the Mean Integrated Square Error (MISE), which consists of two
parts, systematic error and noise. A careful analysis shows that in
the standard PIC methods noise is the dominate error, and the noise
level can be reduced if we select different shape functions that are
capable of balancing the systematic error and the noise. To improve performance,
we use the von Mises distribution as the shape function and seek an
optimal particle width that minimizes the MISE, represented by a Cross-Validation (CV) function.
This procedure significantly reduces both the noise and the MISE for
PIC simulations. A particle-wise width adjustment algorithm and a
width update algorithm are further developed to reduce the MISE.
Simulations using the examples of Langmuir wave and Landau Damping
demonstrate that the KDE algorithm developed in the present study reduces 
the noise level on density estimation by 98\%, and gives a much more accurate result on the linear
damping rate compared to the standard PIC methods. 
Meanwhile, it is computational efficient that can save 40\% time to achieve the same accuracy.
\end{abstract}
\maketitle

\section{Introduction}

Noise is a major concern in particle-in-cell (PIC) simulations \cite{Dawson1983,hockney1988computer,birdsall2004plasma,Okuda1972,Cohen1982a,Langdon1983,Lee1983,Cohen1989a,Liewer1989,Friedman1991,eastwood1991virtual,Cary1993,villasenor1992rigorous,Parker1993,Grote1998,Decyk1995,Qin00-084401,Qin00-389,Qiang2000,Davidson01-all,Chen03-463,Qin01-477,esirkepov2001exact,Vay2002,nieter2004vorpal,Huang2006,Chen2011,squire2012geometric,xiao2013variational,xiao2015variational,xiao2015explicit,qin2016canonical,he2016hamiltonian,kraus2017gempic,Xiao2018}.
This problem stems from the usage of pseudo-particle. The spirit of
pseudo-particle is to combine a large group of real particles into
a pseudo one in order to reduce the computation cost. However, the
reduction of simulation particles boosts the noise in simulation because
the noise level is inversely proportional to the square root of the
number of particles. To handle this, an appropriate choice of shape function can effectively
reduce the noise. To balance the accuracy
and computational costs, a triangular function shape with a width of double grid
size is used in standard PIC methods,
which is known as the first order weighting scheme \cite{birdsall2004plasma}.
This simple shape function alleviates some noise, meanwhile it leaves
a large room to improve. The kernel density estimation (KDE) algorithm
\cite{friedman2001elements,silverman1986density,li2007nonparametric,fukunaga1975estimation},
which is a nonparametric method developed in machine learning and widely applied
in big data science \cite{dinardo1995labor,peterka2016self}, computer
vision \cite{elgammal2002background}, biology statistics \cite{valouev2008genome,taylor2012validating,vermeesch2012visualisation},
and quantitative finance \cite{li2007nonparametric}, provides a path
forward to find a high cost performance ratio estimator of the density function.

According to the KDE theory, the probability density is estimated
by summing up kernel functions, namely the shape functions, located at
all samples, which are simulation particles in PIC simulations. The
key point is to determine the kernel function, and it can be solved
by two steps. First, the shape of the kernel function is designed.
Second, the size of the kernel function is selected. Setting the shape 
triangular and length of kernel double the length of spatial grid, the KDE
algorithm is the same as the standard PIC method. If the shape and
size of the kernel function are allowed to vary, the KDE method provides
a tool to systematically reduce the noise level of the PIC simulations.
In this paper, we initiate an investigation in this direction.

Of course, different noise reduction techniques had been investigated
in the past. For example, low-pass filters \cite{birdsall2004plasma}
can be used to filter out high frequency components, which may contain
undesirable noise. Another important technique is the delta-f method
\cite{Parker1993,Chen03-463,Qin00-084401,qin2003nonlinear,qin2003nonlinearaps,qin2008weight},
applicable for simulating system with small perturbations. The KDE
algorithm developed in the present study can be viewed as a new attempt
in this endeavor using modern techniques exemplified by machine learning
and big data science. 

We adopt the wrapped normal distribution \cite{gumbel1953circular,oliveira2012plug,taylor2008automatic}
as the shape of kernel function. This function can be closely approximated
by Von Mises distribution \cite{collett1981discriminating}, which
is infinitely differentiable everywhere and satisfies the periodic
boundary conditions. And there are many efficient algorithms
\cite{press2007numerical,harrison2009fast} to compute this function.

The width of kernel function will be optimized according to the KDE
theory \cite{ushakov2012bandwidth,heidenreich2013bandwidth,hardle1990bandwidth,sheather1991reliable}.
The goal is to minimize the Mean Integrated Square Error (MISE) \cite{rosenblatt1956remarks},
which is the integrated square of the difference between the estimated
distribution and the true distribution. The MISE consists of two parts,
the bias and the variance. The former corresponds to the systematic
error, and the latter is the noise for PIC simulations. The optimized width
is the one that minimizes the MISE. This is different from the
standard PIC methods, where one selects a small fixed width, i.e., the double
grid size, which yields a small system error (bias), but a large noise
(variance). Obviously, the KDE algorithm has an advantage, because
it balances the bias and variance. The bias-variance trade-off is
a key issue in all types of supervised machine leaning. It is interesting
to realize that it is also the dominant factor in the selection of
particle shape function and width for PIC algorithms. 

Since the knowledge of true density distribution is always absent in practice,
the MISE cannot be calculated directly. To estimate it, there are two 
approaches\cite{li2007nonparametric,silverman1986density}.
One is the plugin method, which uses a pivot width to obtain
an approximate density for estimating the MISE by a first order expansion.
However, the results involve the second derivative of the density
function in the denominator. In PIC simulations, density perturbations
often contains small, short wavelength structures. The reciprocal
of second derivative of density function can be so large that introduces
significant numerical error. Therefore, the plugin method is not applicable
for simulation purpose.

Instead, we adopt the other method, i.e., the cross-validation (CV)
method \cite{silverman1986density,friedman2001elements,li2007nonparametric}.
The idea is to estimate the true distribution using a leave-one-out
distribution, constructed by summing through all samples expect for
the one at the position of evaluation. With this technique, the MISE
can be estimated up to a constant by which the optimization can then be carried
out.

Some distribution may have sharp peaks or narrow valleys. Because
judgment of the optimal width is evaluated according to an integrated sense,
the estimated distribution may ``cut the peak and fill the valley''\cite{silverman1986density}.
To overcome this difficulty, a variate width adjustment is utilized.
The width is shrunk around high density areas and is enlarged in
low density places. This adjustment will further reduces the MISE,
because it lightens the bias when the density is large and alleviates
the variance when the density is small.

After determining the shape and the criteria optimal width, we further allow
the width to be updated through simulations. For this purpose,
steps must be decided at which the width should be resized. The simplest
approach is to recalculate the width at every time-step. But it suffers
from two critical disadvantages. First, the computational cost is
quite expansive since one has to solve a optimization problem at each
time-step. Second, when the distribution is close to a uniform distribution,
the optimal width will be infinitely. Therefore, an more sophisticated
update algorithm is needed. We will adjust the kernel size only at
the steps at which the density is significantly non-uniform as calibrated
by the Anderson-Darling test\cite{andersondarling1954,razali2011power},
meanwhile the density changed significantly since the previous width update. At
all other time-steps, the kernel size is not changed. 

We note that the idea of KDE has appeared previously in the smoothed
particle hydrodynamics (SPH) method \cite{monaghan1992smoothed,price2012smoothed}
in fluid simulations. In this context, the kernel width is determined
by solving simultaneous equations involving one exogenous parameter, which
is selected according to human experience. Therefore, the SPH method does not
determine an optimal kernel size using a closed systematic approach based
on first principles. Diego and Kai used wavelets as kernels to estimate
the density \cite{del2010wavelet}. However, the order of wavelet
transformation is left to be selected manually. Since the order implicitly decides
the width of wavelet kernels, this estimation does not provide a data-driven
width selection mechanism either. The KDE algorithm provides 
a self-consistent path to determine the kernel width.

As numerical examples, Langmuir wave and its linear Landau damping
are simulated by a PIC method using the KDE algorithm to estimate the density. 
For the Langmuir wave, we compare the MISE of optimal width for the CV function
and standard PIC method to demonstrate the reduction of total error by
the KDE algorithm. The density and electric field
calculated by the KDE algorithm show
a significant reduction in noise level as well, compared with the standard
PIC method. In addition, the relation between the optimal kernel width
and the number of simulation particles is investigated.
A larger number of particles requires a smaller optimal width. For the Landau
damping, simulation results using the common random numbers variance
reduction technique\cite{law2007simulation,platen2010variance} show
that the linear damping rate is improved significantly when the KDE
algorithm is applied. [!!!This place need more summaries!!!]

This paper is organized as follows. Section II introduces the methodology
of the KDE algorithm with four subsections on von Mises distribution,
width selection, sample-wise adaptive method, and width update algorithm.
In Section III., Langmuir wave and its Landau damping are simulated
to demonstrate the noise-reduction effect of the KDE algorithm.

\section{Noise reduction using the KDE algorithm}

\subsection{The von Mises Distribution}

Among the various boundary conditions for plasma simulations,
periodical boundary condition is frequently used \cite{birdsall2004plasma}.
To apply the KDE algorithm to PIC simulations, we select
a kernel that satisfies the periodic boundary condition. The normal
distribution function is a commonly used as a kernel in KDE algorithm
with infinite smoothness \cite{li2007nonparametric}. However, the
normal distribution is defined on the entire real line and does not
satisfy the periodic boundary condition. This conflict is resolved
by using the following wrapped normal distribution \cite{gumbel1953circular}
as the kernel, 
\begin{equation}
f_{WN}(\theta;\mu,\sigma)=\frac{1}{\sigma\sqrt{2\pi}}\sum_{k=-\infty}^{+\infty}\exp\left[\frac{-(\theta-\mu+2\pi k)^{2}}{2\sigma^{2}}\right].
\end{equation}
The wrapped normal distribution is just the sum of all normal distribution
shifted by periods of $2\pi$. To reflect the periodic property
of this distribution, the position is denoted as $\theta$ instead
of $x$. Obviously, it automatically satisfies the periodical boundary
condition.

The parameter $\mu$ controls the peak position of the distribution,
and the parameter $\sigma$ is not the variance but convey a similar concept. A small $\sigma$
results in a sharp peak, and the distribution converges to normal
distribution when $\sigma/2\pi\ll1.$ A large $\sigma$ flattens the
distribution, and the distribution converge to uniform distribution
as $\sigma$ goes to infinity.

Instead of summing up the infinite series directly, the wrapped normal
distribution can be approximated by the von Mises distribution \cite{collett1981discriminating}
defined as 
\begin{equation}
f(\theta|\mu,\kappa)=\frac{e^{\kappa cos(\theta-\mu)}}{2\pi I_{0}(\kappa)},\label{eqn:vonMises}
\end{equation}
where $I_{0}$ is the modified Bessel function of order 0. The parameter
$1/\kappa$, analogous to the $\sigma^{2}$ in the wrapped normal
distribution, measures the concentration. The von Mises function can
be efficiently computed using many algorithms \cite{harrison2009fast},
and we implement the polynomial fit procedure \cite{press2007numerical}.

\subsection{The optimized size using CV optimization}

According to the KDE theory, given $n$ independent samples $\{X_{i}|i=1\ldots n\}$
from identical distribution $f(x)$, the probability density
distribution can be estimated according to the following formula,
\begin{equation}
\hat{f}(x)=\frac{1}{n}\sum_{i=1}^{n}\frac{1}{h}K\left(\frac{X_{i}-x}{h}\right).\label{eqn:kde}
\end{equation}
Here, $K$ is the kernel function, $h$ is the bandwidth, or called the width
of the kernel. A larger width results in a lower and flatter shape
function, and a smaller width gives a higher and narrower shape function.
If the kernel function satisfies 
\begin{gather}
\int K(x)dx=1,\\
K(x)=K(-x),\\
\int x^{2}K(x)dx<\infty,
\end{gather}
the estimation is consistent \cite{silverman1986density}. 

The optimized width that minimizes the error of the density
estimator is sought. Before we dive into the error analysis, we must clarify
how the error of the density estimator is defined. In standard PIC
methods, the accuracy of the density distribution is evaluated on
sampling grid points, as measured by the bias error. 
However, it is not the only error that matters,
because the variance error of the estimation is also important.
If a larger width is chosen, the kernel function will be flatter,
and will give a smoother density estimation. It loses detailed
information of the distribution. If a small
width is chosen, the resulting density estimation will be spikier, and it reflects
more details of the density information. As a price paid, it
introduces more statistical noise in the density estimation, yielding
to a large variance. An optimal density estimator needs to keep a balance
between bias and variance. This is the well-known bias-variance trade-off
problem. 

Quantitatively, the error of a density estimator is defined as the
Mean Integrated Square Error (MISE) \cite{rosenblatt1956remarks},
\begin{equation}
MISE=E\int_{0}^{2\pi}\left[f(x)-\hat{f}(x)\right]^{2}\,dx.
\end{equation}
In this paper, the expectation $E$ is taken on
random variables $\{X_{i}|i=1\ldots n\}$, contained in the definition
of $\hat{f}(x)$. The MISE essentially is the sum of squared bias and
variance, because at each $x$ we can decompose the mean square error
as
\begin{align}
E\left\{ [f(x)-\hat{f}(x)]^{2}\right\}  & =E\left\{ [f(x)-E\hat{f}(x)+E\hat{f}(x)-\hat{f}(x)]^{2}\right\} \\
 & =\left[f(x)-E\hat{f}(x)\right]^{2}+E\left\{ [\hat{f}(x)-E\hat{f}(x)]^{2}\right\} \\
 & =bias^{2}+variance.
\end{align}
The bias part, by definition, indicates the deviation of the estimated
density from the true value, and thus corresponds to a systematic
error. The variance part measures the fluctuation of the density estimation,
which is known as noise for PIC simulations. The MISE is the integral of 
sum of the variance and the square of the bias over all $x$, 
which provides a global measure of estimation error.

The key problem now is to find the minimum of MISE. There are several
approaches. A common and simple way is to use the plugin method in
circular data \cite{oliveira2012plug,taylor2008automatic}. For the
von Mises kernel defined in Eq.~(\ref{eqn:vonMises}), as the sample
number goes to infinity, the variance is asymptotically proportional
to $\kappa^{1/2}$, and the bias is proportional to $\kappa^{-2}$
asymptotically \cite{di2009local,taylor2008automatic,oliveira2012plug}.
The size of kernel decreases as $\kappa$ increases,
the bias will be reduced and the variance will be enlarged. On the
contrary, when the size of kernel increase, the bias will become large
and the variance will be controlled. The multiplication of forth power
of leading term of variance width the bias is a constant, so there
will be an optimized point that minimized the MISE that best balance
the variance and bias error. But this plugin method requires knowledge
on the second derivative of the true distribution, which is hard to
obtain. In addition, for a distribution with small, short-wave-length
perturbation, calculation of the second derivative induces large numerical
error. Instead, we will use the cross-validation
(CV) method \cite{friedman2001elements,silverman1986density,li2007nonparametric}.
The definition of MISE involves the true distribution of density,
which is always unknown. The CV method provides a new pathway to estimate
the MISE.

The MISE can be expanded as 
\begin{align}
MISE & =E\int\left[\hat{f}(x)-f(x)\right]^{2}\,dx\label{eqn:MISE}\\
 & =E\int\hat{f}(x)^{2}\,dx-2E\int\hat{f}(x)f(x)\,dx+\int f(x)^{2}\,dx.\label{eqn:expanded_MISE}
\end{align}
The last term can be viewed as a constant since it depends only on
$f(x)$ and thus independent from the kernel width. Given a width value,
the first term can be calculated directly using Eq.~(\ref{eqn:kde}),
because it only involves the estimated density itself. The second
term is more difficult, since the true density is involved. The second term can
expressed in expectation form as 
\begin{equation}
\int\hat{f}(x)f(x)\,dx=E_{X}[\hat{f}(X)].\label{eqn:secondterm}
\end{equation}
The expectation on the right hand side of Eq.~(\ref{eqn:secondterm})
could be estimated by the average of $\hat{f}(X_{i})$, i.e. $\sum_{i=1}^{n}\hat{f}(X_{i})/n$.
However, this estimation leads to $h=0$ when minimizing the MISE.
This fact is proved in Appendix A. To overcome this difficulty, the
leave-one-out estimator \cite{friedman2001elements} is applied. The
expectation is estimated as 
\begin{equation}
\hat{E}_{X}[\hat{f}(X)]=\frac{1}{n}\sum_{i=1}^{n}\hat{f}_{-i}(X_{i}),\label{eqn:Ehat}
\end{equation}
where 
\begin{equation}
\hat{f}_{-i}(X_{i})=\frac{1}{(n-1)h}\sum_{j\neq i,j=1}^{n}K\left(\frac{X_{i}-X_{j}}{h}\right)
\end{equation}
is the leave-one-out estimator of Eq.~(\ref{eqn:secondterm}).
There also exists a similar jackknife re-sampling
method \cite{miller1974jackknife} to achieve the same goal.

The CV function of kernel width $h$ is calculated by 
\begin{align}
CV(h) & =\int\hat{f}(x)^{2}\,dx-2E[\hat{E}_{X}[\hat{f}(X)]]\\
 & =\frac{1}{n^{2}h^{2}}\sum_{i=1}^{n}\sum_{j=1}^{n}\int K\left(\frac{X_{i}-x}{h}\right)K\left(\frac{X_{j}-x}{h}\right)\,dx\nonumber \\
 & \qquad-\frac{2}{n}\sum_{i=1}^{n}\left[\frac{1}{(n-1)h}\sum_{j\neq i,j=1}^{n}K\left(\frac{X_{i}-X_{j}}{h}\right)\right]\\
 & =\frac{1}{n^{2}h}\sum_{i=1}^{n}\sum_{j=1}^{n}\bar{K}\left(\frac{X_{i}-X_{j}}{h}\right)\nonumber \\
 & \qquad-\frac{2}{n(n-1)h}\sum_{i=1}^{n}\sum_{j\neq i,j=1}^{n}K\left(\frac{X_{i}-X_{j}}{h}\right)\label{eqn:cv_def}
\end{align}
where $\bar{K}(x)=\int K(x-t)K(t)\,dt$ is the two-fold convolution
\cite{li2007nonparametric}. It can be proved that $\bar{K}(x)$
is a symmetric probability density function as well.

Since it can be proved \cite{silverman1986density} that 
\begin{equation}
E[\hat{E}_{X}[\hat{f}(X)]]=E[E_{X}[\hat{f}(X)]],
\end{equation}
in which the outer expectation $E$ is taken on $X_{i}$ and inner
expectation $E_{X}$ is taken on $X$, function $CV(h)+\int f^{2}(x)dx$
is an unbiased estimator for the MISE. To minimize $E[CV(h)]$ is
equivalent to minimize the MISE, since $\int f^{2}(x)dx$ is a constant independent
from $h$. Assuming that $\hat{h}$ that minimizes $CV(h)$ also approximately
minimizes $E[CV(h)]$, $\hat{h}$ is then taken as the optimal width for
density estimation. Thus, the CV function can be used as the proxy
of the MISE, and we can apply various algorithms, such as Newton gradient
method, to minimize the CV function. The KDE algorithm replaces the
fixed double grid width in the standard PIC methods with an optimal
width that balances the variance error (noise) and the bias error
(systematic error).

\subsection{Adaptive width}

By minimizing of the CV function, an optimal width is obtained.
However, if the density has sharp peaks or narrow valleys, it is more likely
that the density estimator that minimizes the mean and integrated
error will ``cut the peaks and fill the valleys''\cite{silverman1986density}.

To avoid this shortcoming, instead of using a constant width for
all samples, the sample-wise adaptive widths will be used. The
adaptive method generates different widths for all sample points,
namely the particles, according to the priori density distribution of the
CV-optimal width. It shrinks the widths for particles where the density
is relatively high and enlarges the widths where the density is relatively
low. This procedure can be intuitively understood as follows. When
the priori density is high, there are more particles crowded, and
the bias can be further reduced with smaller width because the 
bias error is the domain error. On the contrary,
in the places where the priori density is low, 
the the variance error dominates, and it needs more particles
to be included to reduce the variance of the estimation. Hence the
width should be enlarged to include more particles in estimation.
With this additional procedure, the MISE is further reduced.

We adopt the adaptive method described by Silverman \cite{silverman1986density}
to perform the adjustment. It is implemented with the following steps.
\begin{enumerate}
\item Find a pilot estimate $\tilde{f}(x)$ by minimizing the CV function. 
\item Define local width factor $\lambda_{i}$ by 
\begin{equation}
\lambda_{i}=\{f(X_{i})/g\}^{-\alpha},\label{eqn:adaptive_width}
\end{equation}
where $g$ is the geometric mean of the $\tilde{f}(x)$, i.e., 
\begin{equation}
\log(g)=\frac{1}{n}\sum\log\left(\tilde{f}(X_{i})\right).
\end{equation}
Here, $\alpha$ is a sensitive factor. Larger $\alpha$ corresponds
to more sensitive dependence of the corrected density on the priori
density. In general, $\alpha$ is chosen to be $0.5$, which keeps
the same sensitivity relative to the priori density and to the samples.
\item Use the adjusted width to estimate the density function by summing
up all kernels located on the samples, 
\begin{equation}
\hat{f}(t)=\frac{1}{n}\sum_{i=1}^{n}\frac{1}{h\lambda_{i}}K\left(\frac{t-X_{i}}{h\lambda_{i}}\right).
\end{equation}
\end{enumerate}

\subsection{Width update strategy}

Contrary to the density estimation scheme discussed above in which 
only a single estimation is necessary, density estimation in plasma
simulation is a dynamical process. The density of particles
is required in each step. Naively, we may wish to optimize the width
of kernels for each step and estimate the density using it. 
But this brings two difficulties. 
It is expansive to solve the optimization. What is even worse,
when the distribution is close to uniform, the optimal width is as
large as the positive infinity, and the width should not be solved
in this situation. 

Therefore, we design an update algorithm to determine at which
steps the width should be re-evaluated. The update algorithm developed
in the present study utilizes a nonparametric test, the Anderson-Darling
test \cite{andersondarling1954}, to tell whether samples obey uniform 
distribution or not. The confidence level of the
test is measured by p-value. Smaller p-value indicates a more
sever deviation from uniform. If we have 99\% confidence level, namely $p\leq0.01$,
on that the distribution is significantly different from a uniform,
then we are able to perform the optimization. For the steps in
which we cannot reject the hypothesis that the distribution is uniform,
the width will not be updated. In order to effectively reduce the
computational cost, the same width is used for successive time-steps
until the change of distribution has significantly accumulated. To
measure the level of change of the distribution, we use the difference between
the p-values of the Anderson-Darling test at the two time-steps.
In other words, we use the difference
of deviation level from the uniform distribution as a measurement
for the changes. Although the same p-value may correspond to different
distributions, this situation happens only when the distribution experiences
a sudden change. Since particle density varies continuously with time,
the variation of the p-value is an adequate indicator for the change
of the distribution. Only when the p-value
at the a time-step reduces by half, we perform the CV optimization.

To sum up, we re-evaluate the width only when the distribution is
different from the distribution of last optimization. 
The update algorithm can be summarized
as follows.

\begin{algorithmic} \If {$\text{p-value}>2\times \text{threshold}$} \State
{$\text{p-value}_{th}=2\times \text{threshold}$;} \ElsIf {$\text{p-value}<0.5\times \text{p-value}_{th}$}
\State calculate $h_{opt}^{target}$ by minimizing $CV$; \State
$\text{p-value}_{th}=\text{p-value}$; \EndIf \State $h_{opt}=\text{adjust-rate}\times h_{opt}^{target}+(1-\text{adjust-rate})\times h_{opt}$
\end{algorithmic}

Here, the adjust-rate variable provides a smooth transition
from the old width to the target optimal width. Empirically, it
is chosen to be 0.05.

\section{Simulations of electrostatic Waves}

\subsection{The Langmuir Waves}

Langmuir wave is a rapid electrostatic oscillation in plasma driven
by Coulomb force. We compare the results of simulations of Langmuir
wave using the standard PIC method and the KDE algorithm.

The equation of motion for electrons is
\begin{align}
m\frac{dv}{dt} &= -eE,\\
\frac{dx}{dt} & = v.
\end{align}
It is discretized with the leap frog method,
\begin{align}
m\frac{v^n-v^{n-1}}{\Delta t}&=-eE_{n-1},\\
\frac{x^n-x^{n-1}}{\Delta t}&=v^n.
\end{align}
The electric field $E$ is determined by Poisson equation
\begin{equation}
    \nabla^2\phi=-\frac{\rho}{\epsilon_0}\quad or \quad \frac{\partial^2\rho}{\partial x^2}=-\frac{\rho}{\epsilon_0},
\end{equation}
which is solved using the following finite difference method on the grids,
\begin{gather}
    E_j=\frac{\phi_{j-1}-\phi_{j+1}}{2\Delta x},\\
    \frac{\phi_{j-1}-2\phi_j+\phi_{j+1}}{\Delta x^2}=-\frac{\rho_j}{\epsilon_0}.
\end{gather}

The density $\rho$ on the grid is determined using 
the KDE method. The leap frog scheme has a second order accuracy on time, and 
the density estimation converges to true density with rate $\log n/nh$ under $L_2$ norm\cite{jiang2017uniform}.

We follows the convention of Birdsall \cite{birdsall2004plasma} to
normalize the physical quantities. The electron plasma oscillation
frequency $\omega_{pe}$ is normalized to 1, so determined the normalized 
density by $\omega_{pe}^2=\rho e^2/m_e\epsilon_0$. The 1D spatial domain
is $[0,2\pi]$ with 512 grid points, and time-step $\Delta t$ is
chosen to be 0.01. We use $2^{14}$ particles in the simulations.
The initial equilibrium density distribution $\rho_{0}$ is uniform
and the perturbation is $\rho_{1}=0.02\cos(x)$. Temperature is set
to zero. We run 10000 time-steps in the simulation and save the sampling
particles' positions and velocities at every time-step.

\begin{figure}
\includegraphics[width=0.5\textwidth]{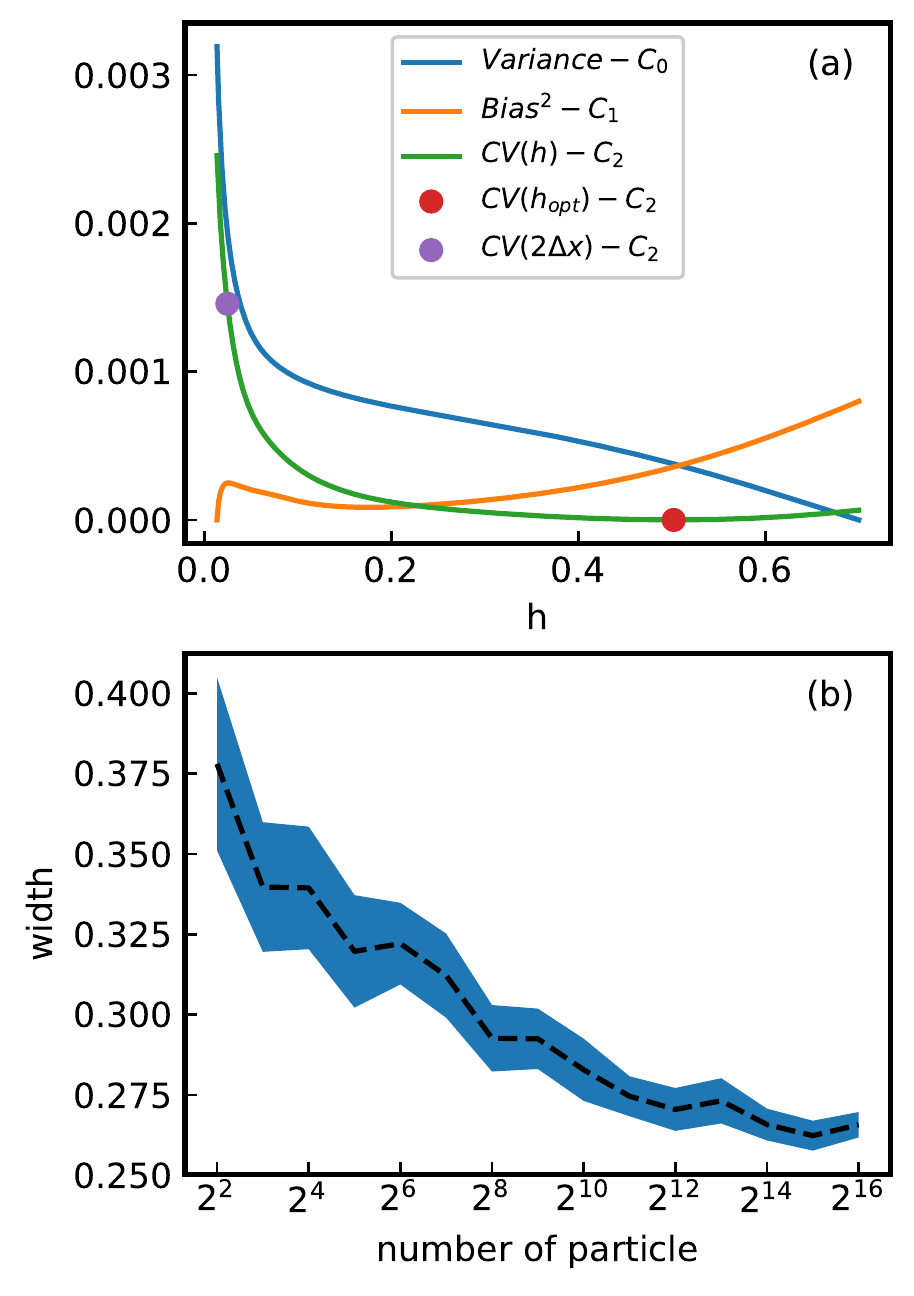} \caption{\label{fig:CV_function}(a) The decomposition of the cross validation
function. The blue curve and the orange curve indicate the variance
and the bias squared term, respectively. The green curve is the total
cross validation function. The purple dot and the red dot denote the
$2\,dx$ for the standard PIC methods and the optimal width, respectively.
(b) The relation between width and number of particles.The blue band
is the 95\% confidence interval with 20 ensembles and the dashed black
curve is the mean. }
\end{figure}

To demonstrate the MISE reduction effect of the KDE algorithm, the
CV functions on given samples versus kernel widths is drawn in Fig.~\ref{fig:CV_function}a.
By subtracting$\int[E\hat{f}(x)]^{2}dx$ from the first term in Eq.~(\ref{eqn:cv_def}),
the variance term is recovered up to a constant that is independent
from h. And similarly the bias squared term, up to a constant, can
be constructed by subtracting the second term in Eq.~(\ref{eqn:cv_def})
from$\int[E\hat{f}(x)]^{2}dx$. The blue curve is the variance, which
decreases with width $h$. The orange curve is the bias squared, which
increases with width when $h$ is greater than 0.1. Summing up the
two curves, we obtained the total CV function (green curve) which
drops dramatically at first, keeps a steady plateau for while, and
then rises up slowly with increasing width. For the convenience of
comparison, these three curves are shifted so that the minimum of
the three curves are all zeros. The purple dot indicates the CV at $h=2\,dx$,
the width used in standard PIC methods, and the red dot is the optimal
point using the CV optimization. It is evident that at the purple
dot, the dominant error is the variance, which reflects the
well-known fact that the noise level in the standard PIC methods is
large. More importantly, Fig.~\ref{fig:CV_function}a suggests that
variance error, or the noise, can be significantly reduced with a
larger width. The MISE error is minimized at the red dot, which is
what is used in our simulations with the KDE aglorithm.

According to the KDE algorithm, the optimal width decreases with the
particle number. When the number is large, the bias error dominates.
When more samples are added, the bias can be compensated by the additional
information carried by the extra samples. To reduce the bias error,
the width calculated by the KDE algorithm will decrease. We demonstrate
the relation between the width and the particle numbers in Fig.~\ref{fig:CV_function}b.
It is clear that the mean of the width as indicated by the dashed
black curve decreases with the number of particles increasing. The 95\% confidence
interval of t-distribution using 20 ensembles for the optimized width
is plotted as the blue band , which shows the same trend. 

\begin{figure}
\includegraphics[width=0.5\textwidth]{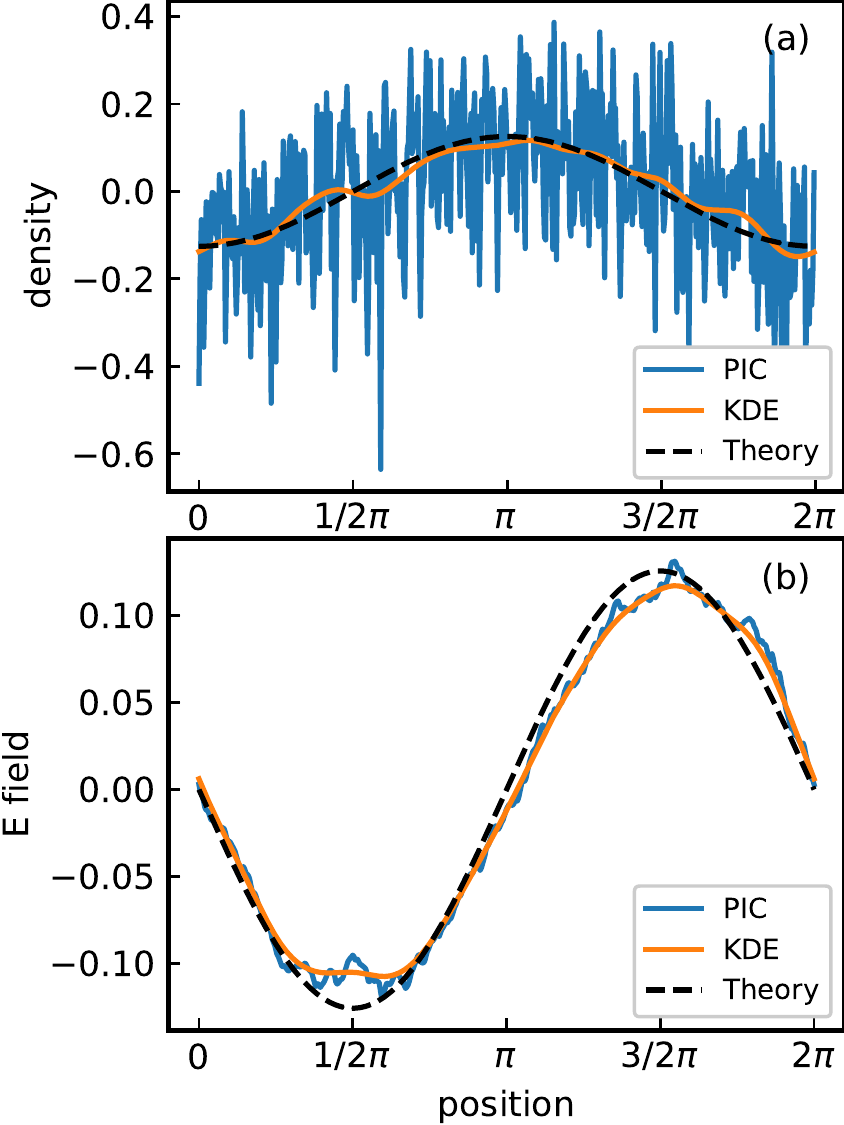}
\caption{\label{fig:density_and_field}The density (a) estimated using the
KDE algorithm (orange curve) and the standard PIC method (the blue
curve), respectively. The black dashed curve is the theoretical density.
The electric fields (b) calculated by the KDE algorithm (orange curve)
and the standard PIC method (the blue curve), respectively. The black
dashed curve is the theoretical electrical field.}
\end{figure}

To further reduce the MISE, adaptive
width for each particle is calculated using the algorithms
described in Section II.C. The density and electrical field at the
initial time-step are shown in Fig.~\ref{fig:density_and_field}.
The black dashed curve in Fig.~\ref{fig:density_and_field}a indicates
the theoretical density function. The blue curve and the orange
curve is estimated the density function using the standard PIC method
and the KDE algorithm, respectively. The standard PIC method suffers
from a large variance error.
On the contrary, the KDE algorithm balances the variance and the bias
error and results in a smooth curve near the the theoretical curve.
The dashed black curve in Fig.~\ref{fig:density_and_field}b is the
theoretical field. The blue curve and orange curve are the field
obtained using the standard PIC method and the KDE algorithm, respectively.
Even though the density is integrated to calculate the field, the
standard PIC method still suffers from a large noise compared with
the KDE algorithm using optimal width.

\begin{figure}
\includegraphics[width=0.5\textwidth]{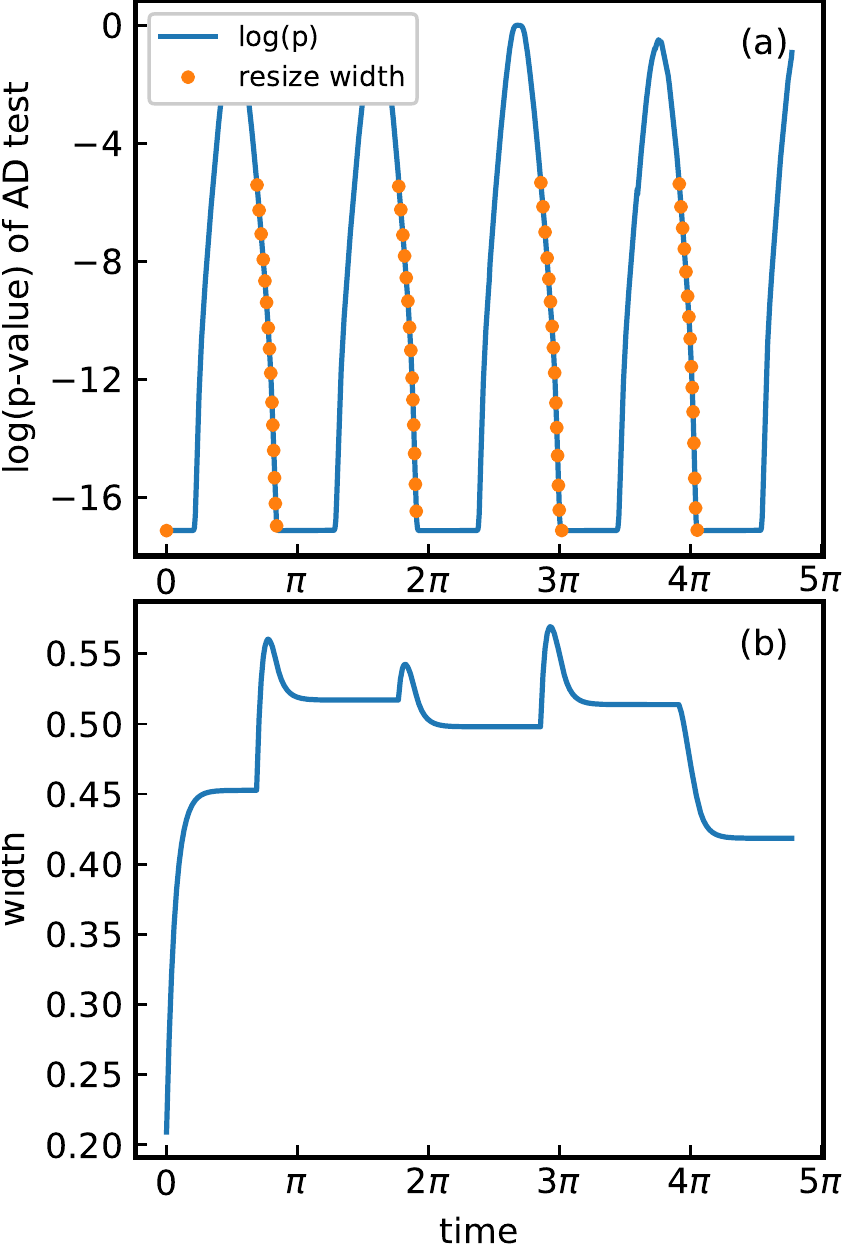}
\caption{\label{fig:update_algorithm}(a) The dynamics of log(p-values) of
the Anderson-Darling test. The dots indicate the time-steps at which
the optimal width is re-evaluated. (b) The width as a function of
time according to the update algorithm. }
\end{figure}

The width is dynamically determined according to update algorithm
described in Sec. II.D. The p-values of Anderson-Darling test is plotted
in Fig.~\ref{fig:update_algorithm}a. The width is updated at the
time-steps in which p-values are less than the threshold and p-values
decreases by half. Orange dots indicate the time-steps at which the
widths are re-evaluated. In Fig.~\ref{fig:update_algorithm}b, the
width is plotted as a function of time.

\begin{figure}
\includegraphics[width=1\textwidth]{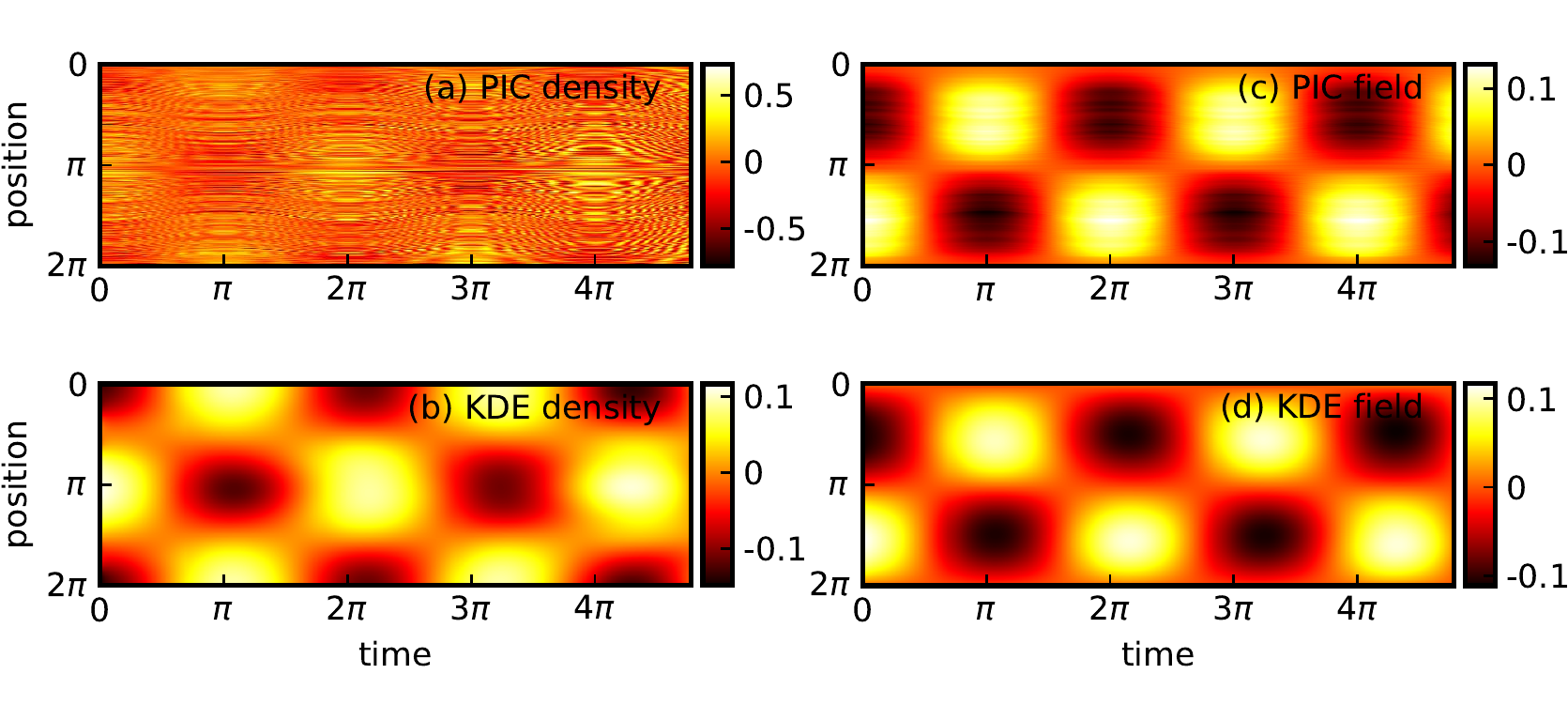} \caption{\label{fig:density_time_space}The density fluctuation of cold Langmuir wave as
a function of time calculated by the standard PIC method (a) and the KDE
algorithm (b). The electric field as a function of time calculated by the
standard PIC method (c) and the KDE algorithm (d).}
\end{figure}

Using the KDE algorithm, the Langmuir wave is simulated. The
density and electric field at all time-steps are compared with those
obtained using the standard PIC method in Fig.~\ref{fig:density_time_space}.

For this specific numerical example, the MISE can be calculated by
definition according to Eq.~(\ref{eqn:MISE}), since the theoretical
density is known. The MISE for density estimation using the standard
PIC method is 0.1260, and that using the KDE algorithm is 0.0023.
It is reduced by 98\% using the KDE algorithm. Using the KDE algorithm leads 
to a 10\% decrease in term of MISE from 0.112 to 0.101 compared to the standard
PIC method.

\begin{figure}
    \includegraphics[width=0.8\textwidth]{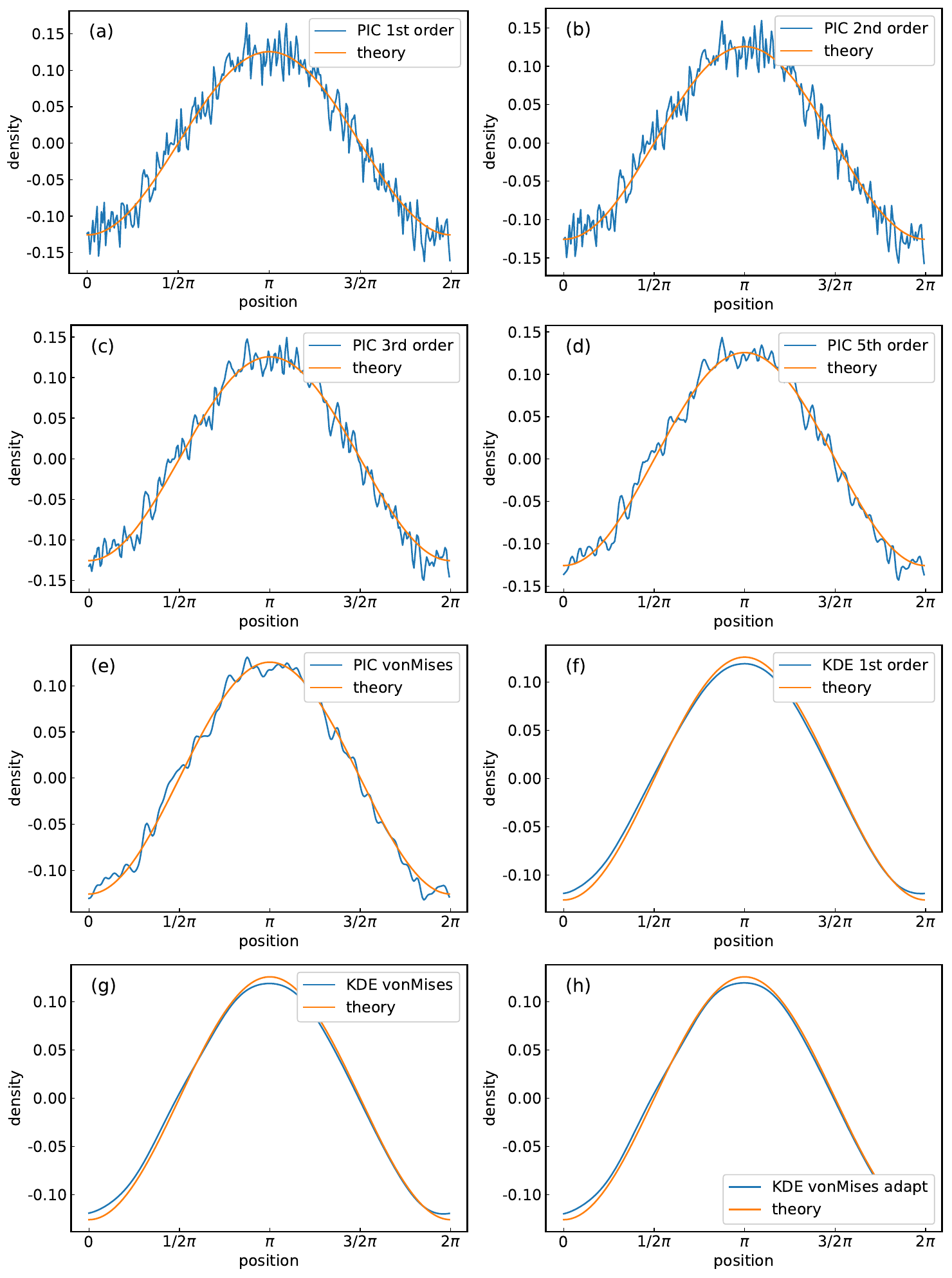} \caption{\label{fig:compare_high_order}
    Noise reduction to different shape functions and kernel widths. In (a-d), the first, second, third and 
    fifth order shape functions with width $(\text{order}+1)dx$ as in standard PIC method, 
    (e) mixture of the von Mises shape with double grid width width as standard deviation, (f) mixture of the 
    first order shape with optimized kernel width, (g) von Mises shape and optimized 
    kernel width by the KDE method, and (h) the von Mises shape and optimal width by 
    the adaptive KDE method.The Blue lines are the density estimations using each 
    method and the orange lines are the theoretical results.}
\end{figure}

The three noise reduction techniques, i.e., high order shape functions, 
kernel width optimization, and adaptive width adjustment can be applied 
combinatorially or individually. 
We have compared the noise reduction effects for different shape functions 
with and without kernel width optimizations using $2^{19}$ particles. The results 
are plotted in Fig.~\ref{fig:compare_high_order}. In (a-d), for the first, second, third and fifth order 
shape functions with width $(order +1)\cdot dx$ as in standard PIC methods, the MISEs 
are $2.28031\times 10^{-3}$, $1.90144\times 10^{-3}$, $1.14667\times 10^{-3}$, and $0.74505\times 10^{-3}$ respectively.
In (e), the von Mises shape is used with double grid width, and the MISE is 
$0.43698\times 10^{-3}$. In (f), by combining the first order shape with an optimized 
kernel width, the MISE is reduced to $0.17233\times 10^{-3}$. In (g), the von Mises 
shape with the optimized kernel width by the KDE method can reduce the MISE 
to $0.15423\times 10^{-3}$. And in (h), the von Mises shape with the optimal width by the 
adaptive KDE method further reduces the MISE to $0.13063\times 10^{-3}$.

From these results we draw the following conclusions. Given the width of shape 
being $(\text{order} +1)\cdot dx$, the MISE decreases as the order of shape increases. The von 
Mises function is an infinitely differentiable function and can be viewed as an 
infinite order shape function. Hence it is the best estimator among all high order 
shape functions and reduces the MISE by 80.83\%. But when using the optimal width, 
even the first order shape can reduce the MISE by 92.44\%, outperforming all previous 
PIC methods. The noise reduction effects due to high order of shape 
functions and optimal kernel width are different. High order shape functions are 
aimed at reducing the bias. On the other hand, the optimized width method takes the 
variance into consideration and provides a smoother and more accurate density estimation. 
The combination of the von Mises shape function and optimal width can reduce the MISE 
by 93.23\%. Finally, the adjusted width can further reduce MISE by 94.27\%. The adjusted 
width effect will be significant when distribution functions contain sharp peaks and 
narrow valleys.

Performance of the three methods is also investigated from the viewpoint 
of computational cost. For comparison, we vary the number of simulation 
particles for different methods to achieve the same accuracy goal, which 
is chosen to be having an MISE less than $2.000\times 10^{-3}$. The 
computational time for different methods is measured and compared.

It is found that the standard PIC method requires $2^{19}$ particles to achieve 
an MISE of $1.997\times 10^{-3}$. The computation costs 
Central Processing Unit(CPU) time of 135\,ms. The KDE 
method with the first order shape function requires only $2^{12}$ particles and 
MISE is $1.977\times 10^{-3}$. It runs for 81.3\,ms. The KDE method with von 
Mises shape function requires $2^{12}$ particles as well and MISE is $1.949\times 10^{-3}$. 
But it needs 845\,ms. According this result, the optimal strategy 
to achieve a given accuracy is to adopt the first order shape function and use 
the KDE to choose an optimal kernel width. Since the optimization to determine 
the optimal width is a fixed cost, it is more preferred to use KDE method in 
long-term simulations. This can then be neglected when averaged over 
each single time-step. The bottleneck using KDE method in parallel computation
with Message Passing Interface(MPI) may hide in communication between processes
since larger kernel length requires more data to exchange.

\subsection{Linear Landau damping rate}

When the electrons have a finite temperature, the Langmuir wave will
be damped by the wave-particle interaction. This is the Landau damping
\cite{stix1992waves}. To simulate this phenomena, particles are drawn
from an initial distribution which is perturbed in configuration space
and Maxwellian with variance $\sigma_{v}^{2}$ in velocity space.
One system consists of initial positions and velocities is call a ensemble.
To do a fair comparison, the two methods should start from the same ensemble.
With the same ensemble, we carry out simulations using the
KDE algorithm and the standard PIC methods respectively. 

In the simulation, the number of grid point is 256, the simulation
time-step is 0.01, and $2^{15}$ particles are used. The initial velocity
distribution is $f(v)=1/\sqrt{2\pi v_{th}^{2}}\exp[-v^{2}/(2v_{th}^{2})]$,
where $v_{th}$ is the thermal velocity. The thermal velocity is normalized
to $v_{th}=0.40$, which determines the spatial normalization. The
initial density of electrons is uniform
modulated by perturbation of the form $\rho_{1}=0.02\,\cos(1.0\,x)$.
This simulation runs for 1500 steps.

\begin{figure}
\includegraphics[width=1\textwidth]{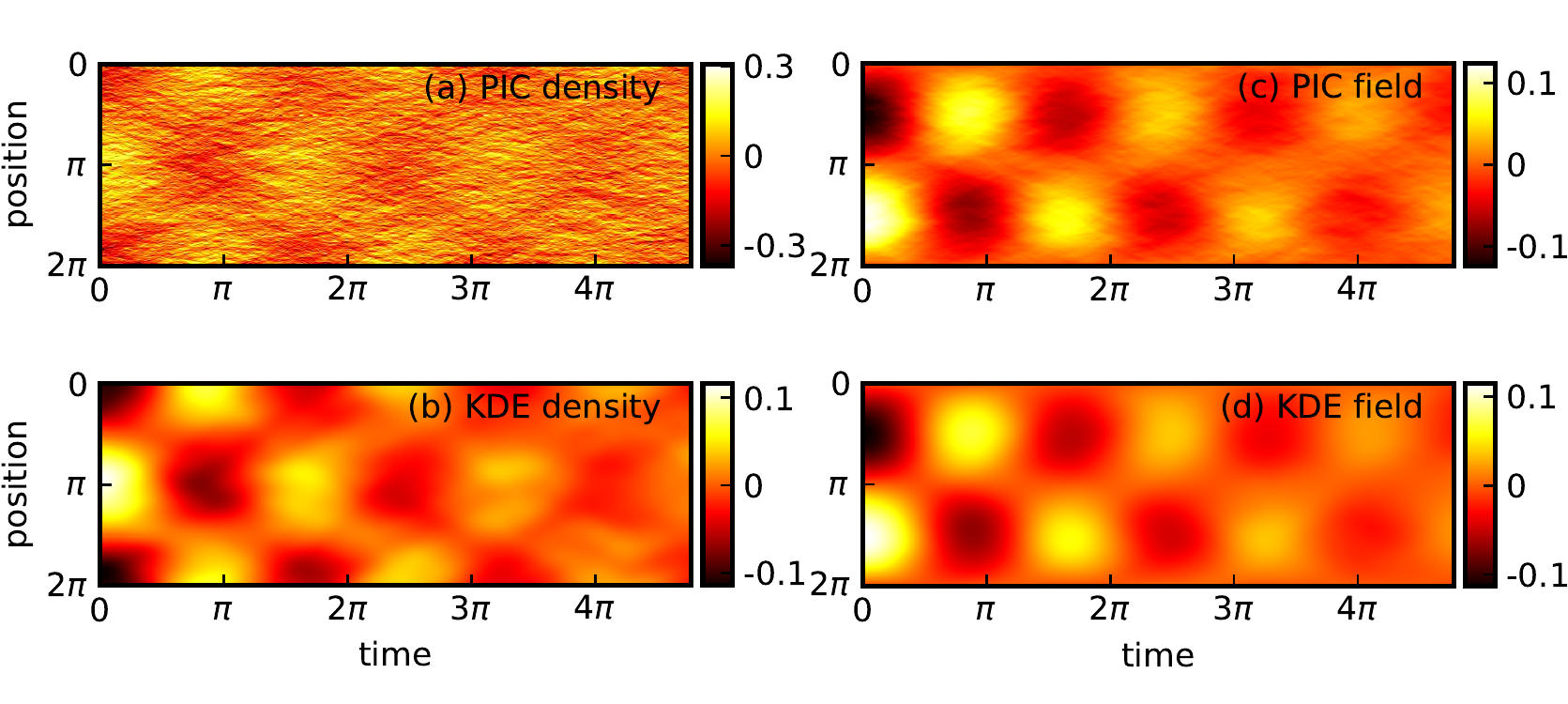} \caption{\label{fig:landau_time_space}The density fluctuation of Landau damping as a function
of time calculated by the standard PIC method (a) and the KDE algorithm
(b). The electric field calculated by the standard PIC method (c)
and the KDE algorithm (d). The normalized thermal velocity is $v_{th}=0.40$.}
\end{figure}

For one specific ensemble, the density distribution with respect to
time using the standard PIC method is plotted in Fig.~\ref{fig:landau_time_space}a
and result using the KDE algorithm is plotted in Fig.~\ref{fig:landau_time_space}b.
The electric field using the standard PIC method and the KDE algorithm
are plotted in Fig.~\ref{fig:landau_time_space}c and Fig.~\ref{fig:landau_time_space}d,
respectively. Compared with the standard PIC method, the KDE algorithm
reduces the noise level on both the density and electrical field.
This noise reduction is expected to render more accurate physical
results. We now show that KDE algorithm generates a more accurate
Landau damping rate.

\begin{figure}
\includegraphics[width=0.5\textwidth]{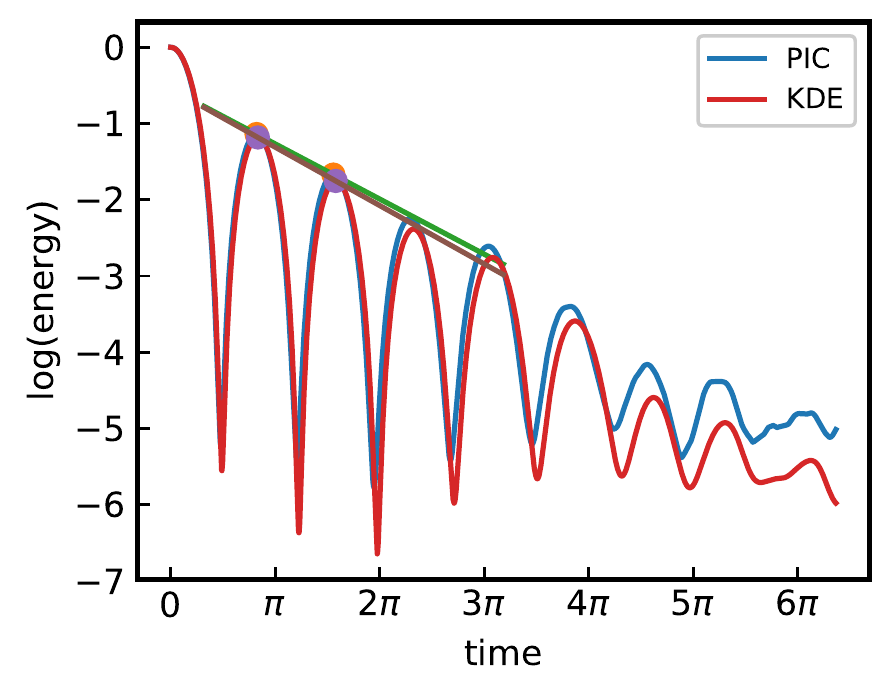} \caption{\label{fig:landau_damping_energy}The logarithm of electric field
energy as a function of time. The result using the standard PIC method
(blue curve) is compared to that using the KDE algorithm (red curve). }
\end{figure}

The Landau damping rate is calculated by the drop of energy of electric
field. When the logarithm of electric energy is plotted, the slope
of the line pass through the first two peaks after $t=0$ is assigned
to be the damping rate. The logarithms of total electric field energy
of this ensemble using the standard PIC method and the KDE algorithm
are plotted in Fig.~\ref{fig:landau_damping_energy} with blue and
red curves respectively. The energy peaks are marked by orange and
purple dots. In this ensemble, theoretical damping rate is -0.096.
Using the standard PIC method, the damping rate is -0.080, and the
relative error is 16.4\%. By contrast, using the KDE algorithm, the
damping rate is -0.087, and the relative error is 9.4\%.

However, comparison of the two methods on only one ensemble is not
enough. To make the comparison more meaningful in statistical sense,
the standard PIC method and the KDE method should be compared on the
mean damping rate averaged over different ensembles corresponding
to the same macroscopic state specified by plasma density, plasma
temperature and velocity distribution. With the common random numbers
variance reduction technique, we draw 30 different ensembles
from the same distribution for both methods.
The absolute error of the mean damping rate using the standard PIC
method is 0.0113 with standard deviation of 0.0032, and the t-statistic
is $t_{PIC}=3.82$. The absolute error of mean using the KDE algorithm
is 0.0038 with standard deviation of 0.0034, and the t-static is $t_{KDE}=1.17$.
At 95\% confidence level, we cannot reject the null hypothesis that
the damping rate obtained using the KDE algorithm is identical to
the theoretical one. But the null hypothesis for the standard PIC
method can be reject, that is to say, the damping rate from the standard
PIC method deviates from theoretical result at 95\% confidence level.
Since the variace is reduced by the common random numbers technique,
the null hypothesis that the error of the standard PIC method is the
same as that of the KDE algorithm can be safely rejected at 95\% confidence
level. Therefore, we can draw the conclusion that the KDE algorithm
improves the simulation result of the linear damping rate by reducing
the noise level.

\begin{figure}
\includegraphics[width=0.5\textwidth]{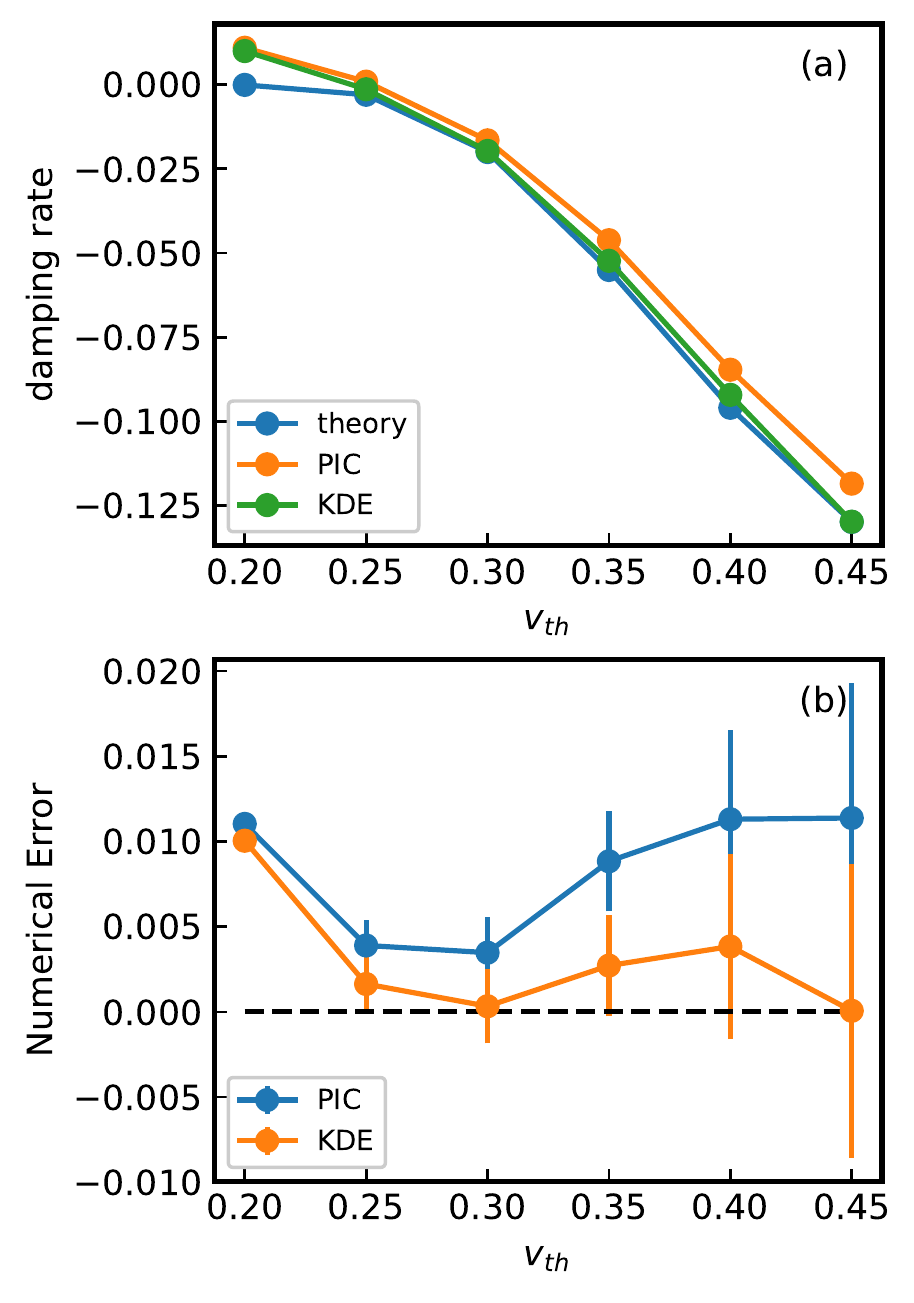} \caption{\label{fig:landau_different_vth}For the different thermal velocity,
the simulations are run for 30 ensembles and the main damping rate
is evaluated. The result using the KDE algorithm (green curve) is
closer to the theoretical damping rate (blue curve) than that using
the standard PIC method (orange curve) (a). The estimation error using
the KDE algorithm (orange curve) and the standard PIC method (blue
curve) are compared in (b). The 95\% confidence intervals are plotted
as error bar.}
\end{figure}

Next, we extend the results to different temperatures. We scan the
thermal velocities from 0.20 to 0.45 for each step of 0.05. The results
are plotted in Fig.~\ref{fig:landau_different_vth}a. The blue curve
is the theoretical result, the green curve is the result using the
KDE algorithm, and the orange is the that using the standard PIC method.
It clear that KDE algorithm is closer to the theoretical
values at every temperature. The numerical errors with error bar of 95\% confidence interval
is plotted in Fig.~\ref{fig:landau_different_vth}b, from which we
can conclude that for each thermal velocity, the mean damping rate
calculated by the KDE algorithm is more accurate than that calculated
by the standard PIC method. 

\section{Conclusion And Discussion}

In this paper, we proposed to use the Kernel Density Estimation (KDE)
algorithm to reduce the noise for Particle-In-Cell (PIC) simulations.
A framework for quantitatively evaluating and minimizing the error
of density estimation for PIC simulations is established using the
KDE theory. Under this framework, the error on particle density estimation
for PIC algorithms, as measured by the Mean Integrated Square Error
(MISE), consists of two parts, the bias error and the variance error.
The bias error is the systematic error and the variance error is the
familiar noise in PIC simulations. A careful analysis showed that
the error on particle density estimation in the standard PIC methods
is dominated by the variance error, which is consistent with well-known
fact that PIC simulations suffer from large noise. Analysis also suggested
that the noise level and total MISE can be significantly reduced by
increasing the width of the kernel function, i.e., particle shape
function. In the KDE algorithm developed in the present study, an
optimal width is obtained by minimizing the MISE, represented by a
Cross-Validation (CV) function, which is constructed using the leave-one-out
estimation. The algorithm is further improved by adopting a particle-wise
width optimization scheme. For the dynamics of the PIC systems, an
efficient width update algorithm is also developed. To test the KDE algorithm, 
we carried out simulations study of the Langmuir wave and its Landau damping. 
Simulation results showed that relative to the standard PIC methods, 
the KDE algorithm can reduce the noise level on density estimation by 98\%, 
and give a much more accurate result on the linear damping rate.

The next step of the research will mainly focus on the choice of different 
loss functions and optimization techniques. The utility of the MISE 
loss function has clear statistical meaning, i.e., it consists of the bias 
and the variance, moreover, the choice of loss function can be and has been extended 
to a wide range in the field of machine learning. Using the $L_1$ norm will enhance 
locality and improve the robustness for outliers. The cross entropy with leave 
one out estimation is another alternative. Furthermore, a more complex KDE 
method can be constructed to model the joined distribution of positions and 
velocities. The loss functions will not only measure the variance and bias of 
position, but also those of phase space as a whole. It will punish particles 
with outlier velocities as well. New optimization method, such as the 
SGD (stochastic gradient descent) method \cite{friedman2001elements} will be 
explored to accelerate the optimal width algorithm without sacrificing accuracy.

Recently, structure-preserving geometric algorithms for plasma physics have been 
actively studied [28-35]. This new class of algorithms are designed to preserve 
the geometric structure of physical systems and bound globally for all simulation 
time-steps the errors on energy, momentum, etc. We plan to combine the KDE method 
with these the structure-preserving geometric algorithms to future improve the 
performance and accuracy of PIC simulations. 

Though the density error is improved by ~98\% than the standard PIC method, 
the field error is only improved by ~10\%. The field is the integration of 
density function, and thus smoother than the density function. Nevertheless, 
it is valuable to have an effective tool for noise reduction on the density. 
There are physical situations, such as in turbulence transport, when the density 
fluctuation is important. On the other hand, to further improve the accuracy 
for the field, we can adopt a loss function to model the error of the field 
and perform optimization, which will be investigated in future studies.

\begin{acknowledgments}
This research is supported by National Natural Science Foundation
of China (NSFC-11775219 and NSFC-11575186), National Key Research
and Development Program (2016YFA0400600, 2016YFA0400601 and 2016YFA0400602),
and the GeoAlgorithmic Plasma Simulator (GAPS) Project.
\end{acknowledgments}

\bibliographystyle{apsrev4-1}
\bibliography{main}

\appendix

\section{Leave-one-out Necessity}

In this appendix, we prove the fact that $h=0$ minimizes the the
KDE estimator defined by Eq.~(\ref{eqn:kde}). In this case, the
CV function is 
\begin{align}
CV(h) & =\int\hat{f}(x)^{2}\,dx-2E[\hat{E}_{X}[\hat{f}(X)]]\\
 & =\frac{1}{n^{2}h}\sum_{i=1}^{n}\sum_{j=1}\bar{K}\left(\frac{X_{i}-X_{j}}{h}\right)\nonumber \\
 & \qquad-\frac{2}{n^{2}h}\sum_{i=1}^{n}\sum_{j=1}^{n}K\left(\frac{X_{i}-X_{j}}{h}\right).\label{eqn:cv_wrong}
\end{align}
By extracting the terms with $i=j$, the CV function is rearranged
into 
\begin{align}
CV(h) & =\frac{1}{nh}\left(\bar{K}(0)-2K(0)\right)\nonumber \\
 & \quad+\frac{1}{n^{2}h}\sum_{i=1}^{n}\sum_{j\neq i,j=1}^{n}\left[\bar{K}\left(\frac{X_{i}-X_{j}}{h}\right)-2K\left(\frac{X_{i}-X_{j}}{h}\right)\right].
\end{align}

Minimizing the MISE is equivalent to minimizing $E[CV(h)]$. We now
investigate how the expectation of the CV function depends on the
width $h$. First, we look at the expectation of kernel function,
\begin{align}
E\left[K\left(\frac{X_{i}-x}{h}\right)\right] & =\int f(\xi)K\left(\frac{\xi-x}{h}\right)\,d\xi\\
 & =\int f(x+h\eta)K(\eta)hd\eta\\
 & =h\int\left[f(x)+f'(x)h\eta+\frac{1}{2}f''(x)h^{2}\eta^{2}+O(h^{3})\right]K(\eta)\,d\eta\\
 & =h\left[f(x)+\frac{h^{2}\kappa_{2}}{2}f''(x)+\frac{h^{4}\kappa_{4}}{4!}f^{(4)}(x)+O(h^{6})\right],
\end{align}
where $\kappa_{2}=\int\eta^{2}K(\eta)\,d\eta$ and $\kappa_{4}=\int\eta^{4}K(\eta)\,d\eta$
are constants depending only on the kernel.

Now we look at each term in the second part of CV function. Since
$X_{i}$ and $X_{j}$ are independent, the first term is
\begin{align}
E\left[\bar{K}\left(\frac{X_{i}-X_{j}}{h}\right)\right] & =\frac{1}{h}E\int K\left(\frac{X_{i}-x}{h}\right)K\left(\frac{X_{j}-x}{h}\right)\,dx\\
 & =\frac{1}{h}\int E\left[K\left(\frac{X_{i}-x}{h}\right)\right]E\left[K\left(\frac{X_{j}-x}{h}\right)\right]\,dx\\
 & =\frac{1}{h}\int E\left[K\left(\frac{X_{i}-x}{h}\right)\right]^{2}\,dx\\
 & =h\int\left[f(x)+\frac{h^{2}\kappa_{2}}{2}f''(x)+\frac{h^{4}\kappa_{4}}{4!}f^{(4)}(x)\right]^{2}\,dx.
\end{align}
The second term is
\begin{align}
E\left[K\left(\frac{X_{i}-X_{j}}{h}\right)\right] & =\int f(x)dx\int K\left(\frac{\xi-x}{h}\right)f(\xi)d\xi\\
 & =h\int f(x)\left[f(x)+\frac{h^{2}\kappa_{2}}{2}f''(x)+\frac{h^{4}\kappa_{4}}{4!}f^{(4)}(x)\right]\,dx,
\end{align}
Therefore, the leading terms for the CV function are
\begin{equation}
CV(f)=\frac{1}{nh}\left(\bar{K}(0)-2K(0)\right)+h^{4}\frac{\kappa_{2}^{2}(n-1)}{4n}\int\left[f''(x)\right]^{2}\,dx-\frac{n-1}{n}\int f(x)^{2}dx+O(h^{5}).
\end{equation}

Assuming that the kernel function has peak at $x=0$ such that for
all $x$, $K(0)\ge K(x)$, which is the case for the von Mises distribution,
we have 
\begin{align}
\bar{K}(0)-2K(0) & =\int K(x)^{2}\,dx-2K(0)\\
 & \leq\int K(0)K(x)\,dx-2K(0)\\
 & =K(0)-2K(0)<0.
\end{align}
Meanwhile, 
\begin{equation}
\frac{\kappa_{2}^{2}}{4}\int\left[f''(x)\right]^{2}\,dx>0.
\end{equation}
It is thus straightforward to show that the the CV function goes to
$-\infty$ when $h\to0$. This proves the fact that $h=0$ minimizes
the KDE estimator defined by Eq.~(\ref{eqn:kde}).
\end{document}